# Reducing the Aspect Ratio of Contact Holes by In-Situ Low-Angle Cross Sectioning

Uwe Scheithauer

82008 Unterhaching
E-mail: scht.uhg@googlemail.com

## Abstract

Auger analysis of high-aspect ratio contact holes of integrated microelectronic devices is a challenging analytical task. Due to geometrical shadowing the primary electron beam and the energy analyser have not the required direct line of sight to the analysis area simultaneously.

To solve this problem sample preparation is needed to flatten the 3-dimensional geometry. Here the new approach of in-situ low-angle cross sectioning is applied. By this method material gets removed inside the Auger instrument while the sample is sputtered by $Ar^+$ ions at nearly grazing incidence utilizing the edge of a mask, which partly covers the sample. A very shallow bevel with respect to the sample surface is produced. Thus along the bevel contact holes with suitable aspect ratios are available for the Auger analysis.

## 1. Introduction

In-situ low-angle cross sections are fabricated inside an Auger microprobe using the equipped depth profile sputter ion gun. The sample gets partly covered by a mask. Then the sample is sputtered with ions. A cross section is produced in the sample due to the shading of the mask. It has been demonstrated, that the slope of this cross section is considerably shallower than given by the geometrical setup [1]. In-situ low-angle cross sectioning has been proven as a powerful tool for depth resolution optimized depth profiling of polycrystalline thin film systems [2][3]. Additionally, this sample preparation technique can be used for metallographic purposes. Therefore the cross section is imaged by elemental mappings, for instance [4].

Here in-situ low-angle cross sections are used to enable the Auger analysis of the bottoms of high aspect ratio contact holes of microelectronic devices. If an Auger microprobe with a cylindrical mirror analyser (CMA) and a coaxially mounted electron gun is used, only for aspect ratios below ~ 0.7 the primary electron beam and CMA have simultaneously access to the contact hole bottom [5]. For higher aspect ratios Auger electrons, which are emitted from the bottom, are geometrically shaded by the contact hole wall. The concept of aspect ratio reduction using in-situ low-angle cross sectioning was validated by a successful Auger analysis of a contact hole bottom. This procedure has advantages compared with other approaches.

## 2. Instrumentation and Data Evaluation

For the Auger measurements presented here a Physical Electronics PHI 680 Auger microprobe was used [6]. The PHI 680 microprobe, an instrument with a hot field electron emitter, has a lateral resolution of approximately ≤ 30 nm at optimum. A lateral resolution of ~ 50 ... 70 nm is achievable under analytical working conditions using a higher primary current.

The PHI 680 microprobe uses a CMA for the energy analysis of the emitted Auger electrons. The CMA has a concentric acceptance area of 42° ± 6° [7][8]. The primary electron beam is guided coaxially to the centrical axis of the CMA.

The Auger microprobe is equipped with a differentially pumped $Ar^+$ ion sputter gun (model: 04-303). If the electron beam hits the sample parallel to the sample surface normal as used here, the incoming sputter $Ar^+$ ions impinge onto the surface under an angle of 75° relative to the surface normal. The instrument is equipped with a sample transfer system. The samples are mounted on sample holders, which are introduced to the main analysis chamber via a turbo pumped vacuum chamber.

The Auger electron spectroscopy (AES) data evaluation uses the PHI software Multipak 6.1. In case of quantification of measured peak-to-peak heights of differentiated intensity signals Multipak 6.1 uses a simplifying model. It assumes that all detected elements are distributed homogeneously within the analysed volume. This volume is defined by the analysis area and the information depth of the electrons used for the measurement. The information depth is derived from the mean free path of electrons [9]. Using this approach one monolayer on top of a sample quantifies to ~ 10 … 30 at% dependent on the samples details.

## 3. Challenge of Contact Hole Analysis

In modern VLSI microelectronic devices, for instance, the aspect ratio of contact holes, which is defined by the depth-to-width ratio, is usually greater than 1. This causes serious problems when the contact hole bottom has to be analysed using an Auger microprobe due to the geometrical angle between the primary electron beam and electron energy analyser take off axis. It restricts the maximum aspect ratio, for which both devices have simultaneously a direct line of sight to the contact hole bottom. Physical Electronics Auger microprobes are equipped with a CMA with coaxially mounted electron gun. With respect to the rotational symmetry axis of the CMA and the primary electron beam, respectively, in a PHI 680 the CMA has an acceptance angle of 42°± 6° from this axis for 360° around it [6][7][8].

In the Auger microprobe a direct line of sight is necessary from the analysis area to both the incoming electron beam and the CMA energy analyser (see fig. 1). If the fine focused electron beam is positioned in the centre of a cylindrical contact hole, for an aspect ratio of ~ 0.45 the blocking of the Auger electrons starts if the electron beam hits the surface with normal incidence. For an aspect ratio of ~ 0.68 and above the CMA has no longer a direct line of sight to the contact hole bottom. Simulations reported in literature show [5], that starting with this aspect ratio the measured Auger intensity is drastically reduced with increasing aspect ratio. For higher aspect ratios the measured intensity is due to Auger electrons emitted from the contact hole side wall and the top plane in the contact hole surroundings.

Different approaches have been implemented to enable reliable contact hole bottom analysis of high aspect ratio contact holes. Mechanical and abrasive sample preparation methods or electrostatic deflection have been used for this.

Microelectronic Si devices may be cracked mechanically outside the instrument. Since the contact holes are repeated multiple on these devices, it is promising to find a contact hole with an accessible bottom along the crack using the secondary electron imaging of the Auger microprobe. But the remaining side wall of the contact hole complicates the situation. Even if the primary beam is focused on the contact hole bottom, with a certain probability stray electrons, which are reflected elastically or with higher kinetic energy from the analysis area at the bottom, may excite Auger electron emission form the contact hole side wall [10]. Due to this effect one can not judge whether low level contaminations are present on the contact hole bottom or on its side wall. To minimize shading of the remaining contact hole side wall the sample has to be tilted to some degree. In this situation more primary electrons, which are scattered in forward direction, will impinge at the side wall and initiate Auger electrons there.

By mechanical, nearly surface parallel polishing the aspect ratio can be reduced [11]. But due to the lubricant needed for this, a contamination of the sample surface is unavoidable. So this attempt makes sense only, if the information using Auger measurements is obtainable after sputter cleaning of the surface inside the instrument. Another drawback of the mechanical polishing is a reaction of sample materials with the water based lubricant. For instance, a serious oxidation of Al layers was observed [12]. The concept of the nearly surface parallel polishing has been advanced by using focused ion beam sample preparation to remove the material and to reduce the aspect ratio this way [13].

A completely other more academic approach is to use an electrostatic deflection between sample and CMA. This way electrons, which can geometrically leave the contact hole with lower angles, are deflected to match the acceptance angle of the CMA [14].

## 4. Aspect Ratio Reduction Using In-situ Low-angle Cross Sectioning

Fig. 1 illustrates the new concept of an aspect ratio reduction using in-situ low-angle cross sectioning. The sample is partly covered by a mask. Since the contact holes are repeated in arrays over the whole sample, the positioning of the mask is uncritical. Utilizing the edge of this mask the sample is sputtered by $Ar^+$ ions inside the Auger microprobe. The $Ar^+$ ions have an impact angle of ~ 15° relative to the surface. In the shadow of the mask a bevel develops. Due to self-alignment effects the slope of the in-situ low-angle cross section is considerably shallower than given by the geometrical setup [1]. In this bevel area the aspect ratio of the contact holes is reduced to a different extend. Since some sputter deposition from the mask to the sample surface in the sub-monolayer region can not be excluded, a suitable mask material should be chosen. It should have no interference with the expected contact hole bottom composition.

Additionally, the drawing shows the measurement situation using a CMA with a coaxially mounted electron gun. For a high aspect ratio the CMA can not see the Auger electrons from the contact hole bottom (fig. 1, sketch at the right). The electrons are blocked by the side walls. If the aspect ratio is reduced sufficiently, the CMA has access to the contact hole bottom (fig. 1, sketch at the left)

## 5. Auger Measurement of a Contact Hole Bottom After In-situ Low-angle Cross Sectioning

An exemplary measurement demonstrates the validity of this concept using the following sample: On a Si substrate, a $TiSi_2$ layer is covered by a 1 μm thick $SiO_2$ layer. By reactive ion etching (RIE) contact holes with ~ 400 nm diameter were etched through the $SiO_2$ layer down to the $TiSi_2$. The RIE was done by a mixture of $Ar^+$ and fluorocarbons. On the one side the $SiO_2$ was etched physically by the $Ar^+$ ion impact and on the other side chemically by the formation of Si fluorides. During the etching carbon fluorine polymer layers build up at all surfaces. Especially when the etching process reaches the $TiSi_2$ a combination of this carbon fluorine polymer layer and Ti fluorides acts as an etch stop [15][16]. Some cleaning has to be done to remove this contamination layer on top of the $TiSi_2$. The general analytical task is to evaluate the contaminations after the etching by AES measurements and to judge on the post etch cleaning procedure.

An in-situ low-angle cross section of the sample was fabricated. In the bevel area contact holes are present with a convenient aspect ratio where a geometrical access to the contact hole bottom is guaranteed and the bottom itself was not sputtered by $Ar^+$ ion impact. Fig. 2 shows a suitable contact hole. As seen by the impression at the contact hole rim, the $Ar^+$ ions impinge on the sample from the upper left direction. Above the hole the $Ar^+$ ions have no interaction with the sample. No bevel slope flatting effect takes place and therefore the impression opposite to the $Ar^+$ ion impact direction has developed. The rectangle in the image marks the measurement area on the contact hole bottom. Within this area an AES survey scan was measured successfully. It is depicted in Fig. 3. The measured signals were quantified according to the algorithm described in section 2. The $TiSi_2$ is detected by the measurement of Ti and Si signals. The measured O signal gives a hint, that the silicide surface is oxidised to some extend. The amount of O can be estimated to be in the order of one monolayer. No serious C and no F signal were detected in the spectrum. To conclude, the cleaning process after the contact hole etching has successfully removed the etch stop layer. A P contamination at the contact hole bottom was detected. It is due to an enrichment of the P dopant of the $SiO_2$ layer during the etching of the contact hole. If small signal contributions of the measured spectrum are interpreted, a possible material redeposition from the elevated structures or from the side wall to the contact hole bottom cannot be excluded, maybe via the inner surfaces of the instrument [17].

To subsume, this exemplary measurement of a cleaned RIE sample shows, that the concept of in-situ low-angle cross sectioning sample preparation enables the analysis of high aspect ratio contact holes. In this case it allows to judge on the success of post RIE sample cleaning of contact hole bottoms of high aspect ratio contact holes.

## 6. Summary

To enable Auger analysis of high aspect ratio contact holes, some kind of suitable sample preparation is indispensable to reduce the pronounced topography. After this preparation the primary electron beam and the energy analyser have the required direct line of sight to the analysis area at the contact hole bottom simultaneously.

Cracking of the Si microelectronic device outside the instrument only partially removes the contact hole side wall. The remaining side wall shadows the CMA to some extent. Unavoidable contributions to the measured signal due to Auger electron emission from the side wall are expected.

Ex-situ nearly surface parallel mechanical polishing can reduce the aspect ratio. But due to this crude process, the original contact hole bottom surface is contaminated or modified in an unpredictable way. So the mechanical polishing is only an applicable preparation, if the sample is sputter cleaned in the Auger instrument first before any measurement. For this reason the information about the original contact hole bottom surface can never be evaluated. The mechanical polishing can be substituted by FIB processing of the sample. However, this additional, expensive instrumentation is necessary.

In-situ low-angle cross sectioning uses the equipped instrumentation of a commercial Auger microprobe. The sample preparation is done inside the microprobe, which minimizes contaminations. Simply the sample is covered by a mask and it is sputtered by $Ar^+$ ions at nearly grazing incidence utilising the masks edge. In the shadow of the mask a bevel develops, which is flatter than given by the geometrical setup due to self-alignment effects. As demonstrated here, in the bevel area contact holes can be found, which have a suitable aspect ratio. A significant measurement of a contact hole bottom was shown after a post RIE cleaning.

## Acknowledgement

All measurements presented here were done using a PHI 680 installed at Siemens AG, Munich, Germany. I acknowledge the permission of the Siemens AG to use the measurement results here. For many fruitful discussions I would like to express my thanks to my colleagues, in particular to F. Bleyl.

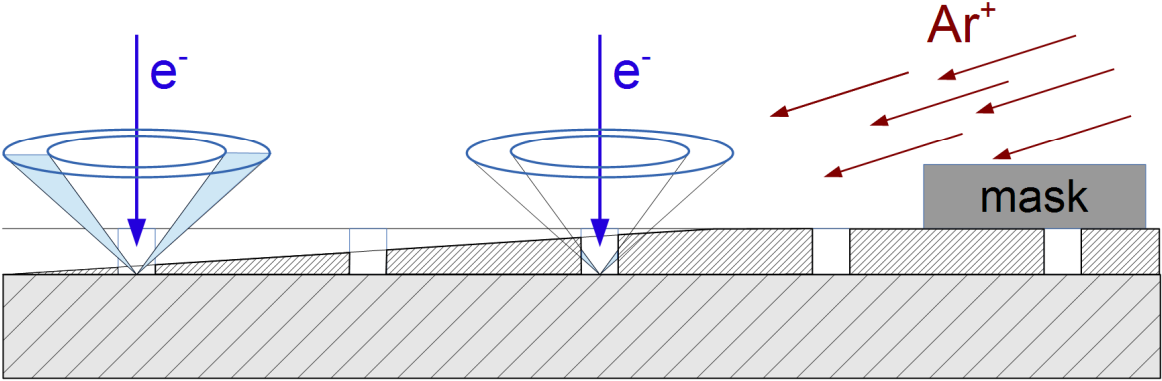

Fig. 1: geometrical situation of a contact hole measurement using an Auger instrument with CMA energy analyser
The aspect ratio is reduced by in-situ low-angle cross sectioning using $Ar^+$ ion sputtering and a mask.

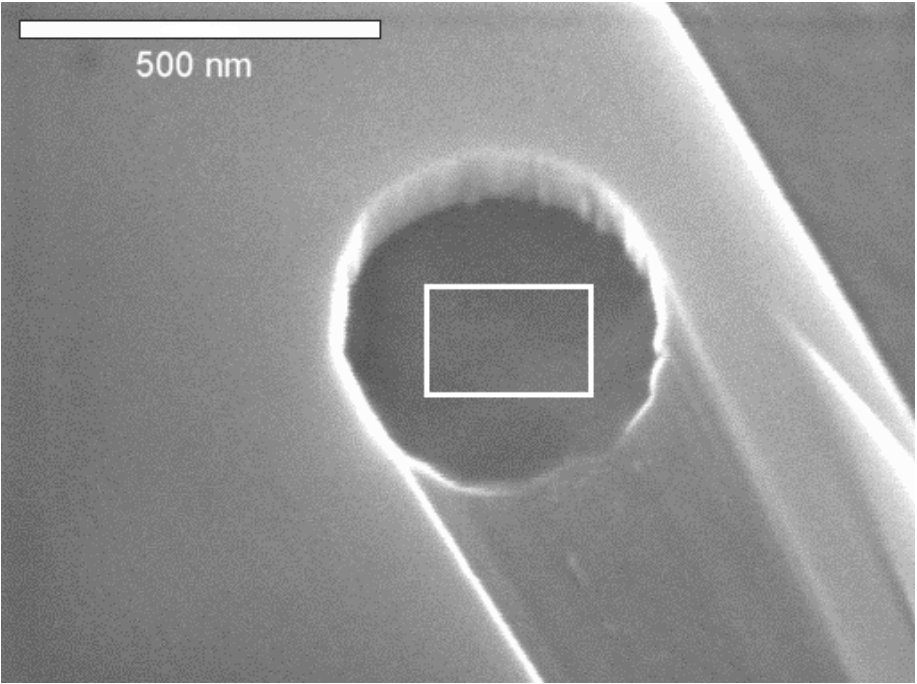

Fig. 2: SEM image of one contact hole after in-situ low-angle cross sectioning

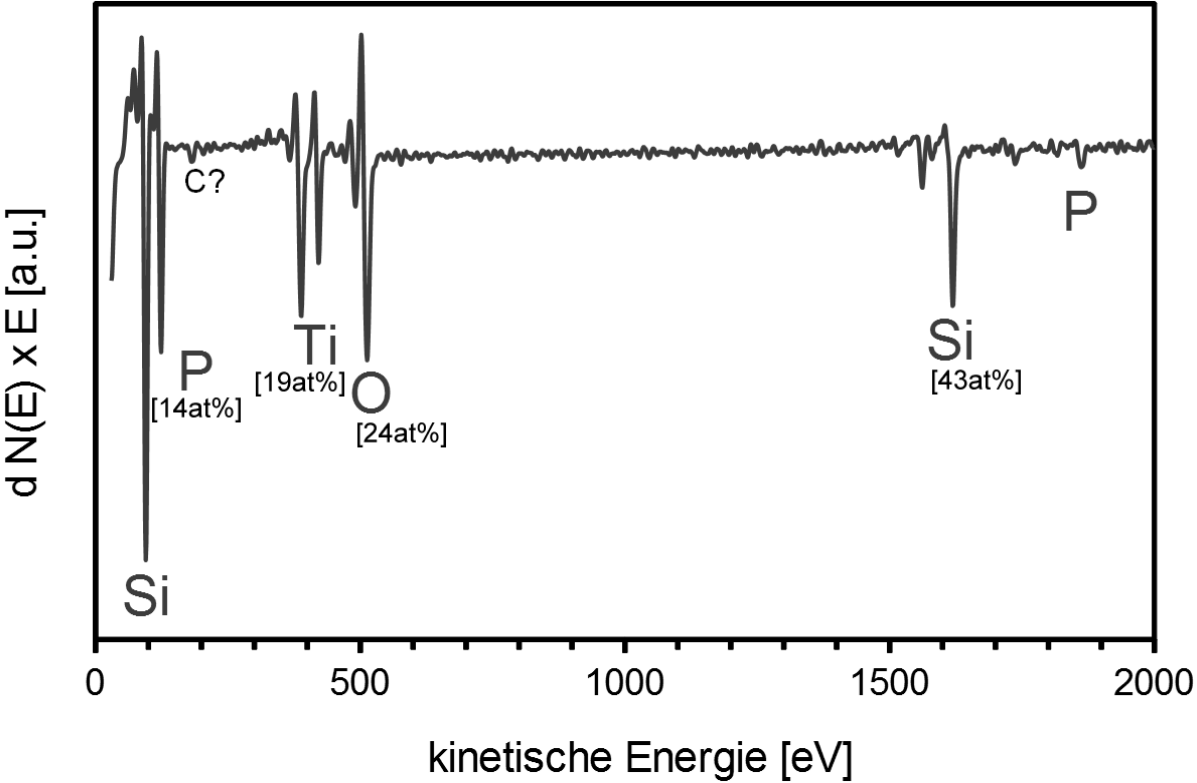

Fig. 3: AES survey scan of the area at the contact hole bottom marked in fig. 2